# JPCam: A 1.2Gpixel camera for the J-PAS survey

K. Taylor[a], A. Marín-Franch[b], R. Laporte[c], F.G.Santoro[d], L. Marrara[e], J. Cepa[f], A. J. Cenarro[b], S. Chueca[b], D. Cristobal-Hornillos[b], A. Ederoclite[b], N. Gruel[b], M. Moles[b], F. Rueda[b], S. Rueda[b], J. Varela[b], A. Yanes[b], N. Benitez[g], R. Dupke[h], A. Fernández-Soto[i], P, Jorden[j], G. Lousberg[k], A. Molino Benito[g], I. Palmer[j], C. Mendes de Oliveira[a], L. Sodré Jr[a]

[a]Universidade de Sao Paulo, IAG, Rua do Matao, 1226, Sao Paulo, 05508-900, Brasil;
[b]Centro de Estudios de Física del Cosmos de Aragón, Plaza San Juan 1, Planta 2, 44001 Teruel, Spain;
[c]Divisao de Astrofísica, Instituto Nacional de Pesquisas Espaciais, Brazil;
[d]New Mexico Tech / MRO - 101 East Rd, Socorro - NM - USA
[e]TopCooler, Av. Dr. Carlos Botelho, 3526 - São Carlos/São Paulo, Brazil;
[f]Instituto de Astrofísica de Canarias, 38200 La Laguna, Tenerife, Spain;
[g]Instituto de Astrofísica de Andalucía (CSIC), Granada, Spain;
[h]Observatorio Nacional, Rua Gal. Jose Cristino, 20921-400, Rio de Janeiro, Brazil;
[i]Instituto de Física de Cantabria (CSIC-UC), E-39005, Santander, Spain;
[j]e2v technologies, 106 Waterhouse Lane, Chelmsford, Essex, UK;
[k]AMOS, Rue des Chasseurs Ardennais 2, B-4031 Angleur, Liège, Belgium;

**ABSTRACT**

JPCam is a 14-CCD mosaic camera, using the new e2v 9k-by-9k 10μm-pixel 16-channel detectors, to be deployed on a dedicated 2.55m wide-field telescope at the OAJ (Observatorio Astrofísico de Javalambre) in Aragon, Spain. The camera is designed to perform a Baryon Acoustic Oscillations (BAO) survey of the northern sky. The J-PAS survey strategy will use 54 relatively narrow-band (~13.8nm) filters equi-spaced between 370 and 920nm plus 3 broad-band filters to achieve unprecedented photometric red-shift accuracies for faint galaxies over ~8000 square degrees of sky. The cryostat, detector mosaic and read electronics is being supplied by e2v under contract to J-PAS while the mechanical structure, housing the shutter and filter assembly, is being designed and constructed by a Brazilian consortium led by INPE (Instituto Nacional de Pesquisas Espaciais). Four sets of 14 filters are placed in the ambient environment, just above the dewar window but directly in line with the detectors, leading to a mosaic having ~10mm gaps between each CCD. The massive 500mm aperture shutter is expected to be supplied by the Argelander-Institut für Astronomie, Bonn.

We will present an overview of JPCam, from the filter configuration through to the CCD mosaic camera. A brief outline of the main J-PAS science projects will be included.

## 1. INTRODUCTION

JPCam is being developed as the primary camera for the Astrofísico de Javalambre[1,2,3,4] (OAJ), a new robotic observatory located at the Sierra de Javalambre (Teruel, Spain) whose primary role will be to conduct all-sky astronomical surveys. The OAJ's main telescope (the JST/T250) is a 2.55m telescope fitted with a 3º diameter wide-field corrector (WFC). The T250 telescope is being supplied by AMOS (Advanced Mechanical and Optical Systems, Belgium) while the WFC is sub-contracted to Tinsley Laboratories (California). The T250 will have a plate scale of

22.67"/mm with an unvignetted 3º field of view (FoV) corresponding to a diameter of 476.4mm; JPCam itself is a 14-CCD close-packed, but not butted, mosaic of wafer-scale CCDs extended over the full FoV.

The goal of the T250 and its panoramic camera is to carry out the so called Javalambre PAU (Physics of the Accelerated Universe) Astrophysical Survey, hereafter J-PAS[5], a photometric all-sky survey of about 8,000 square degrees using 54 narrow band filters of 13.8nm in band-width contiguously spanning a 370 to 920nm wavelength range, plus two broad U and Z band filters[6]. An additional broad band R filter is also foreseen for deep imaging and weak lensing observations. The J-PAS survey will be completed in about 4-5 years from first light and is expected to reach a 3 arcsec aperture magnitude depth of AB= 22.5 − 23.5, depending on the wavelength, at a $5\sigma$ level.

JPCAM has the three main subsystems:

- the non-cryogenic subsystem, mounted directly to the Instrument Support Structure (ISS), which comprises the filter exchange mechanism and shutter working at ambient temperature. In JPCam parlance, this is referred to as the filter shutter unit (or FSU);

- the cryogenic camera subsystem (or Cryo-Cam) comprising the entrance window to the dewar; the focal plane assembly, referred to here as the focal plane cold plate (FPCP), containing the science, wave-front sensors (WFSs) and acquisition and guide sensors (AGs) and their associated controllers; the cooling and vacuum systems and the image acquisition electronics and control software.

- The Cryo-Cam is mounted neither to the FSU nor directly to the ISS but instead is bridged to the ISS via a hexapod actuator system (HAS) which actuates the Cryo-Cam in response to the WFS signals from within the camera itself.

The structural layout allows the considerable weight of the Cryo-Cam to bi-pass the much lighter FSU which does not require to be actuated by the HAS.

## 2. TECHNICAL REQUIREMENTS AND FUNCTIONALITY OF JPCAM

The goal of the system design of JPCam is to maximize the efficiency of field and wavelength coverage in the context of a system which has the 56 filters placed outside the Cryo-Cam dewar; a necessity given the number of filter exchanges required. This in turn implies that the filters are arranged in a set of filter trays where each filter isolates the light onto an individual CCD in the mosaic. Each detector then sees a subset of the 56 filters corresponding to the tray that is in use at the time. It thus becomes essential to maximize the CCD format while allowing for a significant gap between each CCD to avoid filter vignetting. Furthermore, given the required pixel sampling (~0.2"/pixel), detectors with pixels in the neighbourhood of ~10μm are required. Full 6-inch wafer chips are now becoming available with formats up to ~10$k^2$; these are not yet fully (4-side) buttable however they do allow JPCam to maximize its field coverage within the corrected FoV by minimizing, as much as is feasible, the optical distance between the "warm" filters and the "cold" CCD focal plane. Such geometrical considerations lead to a solution using a 14 CCD mosaic with the CCD distribution given as in Fig.1. The FSU is consequently designed to admit 4 filter trays with each containing 14 square intermediate-band filters corresponding to the 14 CCDs of the mosaic. In addition there will be a $5^{th}$ filter tray which will hold 14 identical broad-band filters for deep survey work. Each CCD will view only its corresponding filter with no cross-talk between them.

This design requires the filters to be as close as possible to the dewar window which then requires the shutter unit to be mounted above, but as close as possible, to the filters; a shutter aperture of ~525mm is required. The shutter has to be able to block and unblock the light going to the detector in such a way that the exposure times are uniform at the >99% level across the FoV. For such a large FoV, a homogeneous illumination can be achieved using "two-curtain" shutters. Similar shutters in size and requirements are produced by the Bonn University group for other large format imagers[7].

The science detectors were selected at the beginning of the program of work; e2v CCDs were chosen for their format ($9k^2$, 10μm pixel, 16-channel) and readout noise performance. The 14 CCD mosaic of e2v detectors can be almost completely inscribed by the FoV of the T250 telescope, as exampled in Fig.1. Within the periphery sectors not occupied by the detector mosaic, auxiliary CCDs are installed for guiding, focussing and wave-front sensing. The size of these auxiliary CCDs are maximized, while staying largely within the FoV of the T250 telescope and are required to have approximately the same pixel size as the science CCDs. A wave-front curvature sensing system, based on pairs of intra- and extra-focus sensors, commands the Telescope Control System (TCS) to maintain high fidelity imaging over the whole focal plane. Again e2v detectors were chosen for the guide and WFS chips.

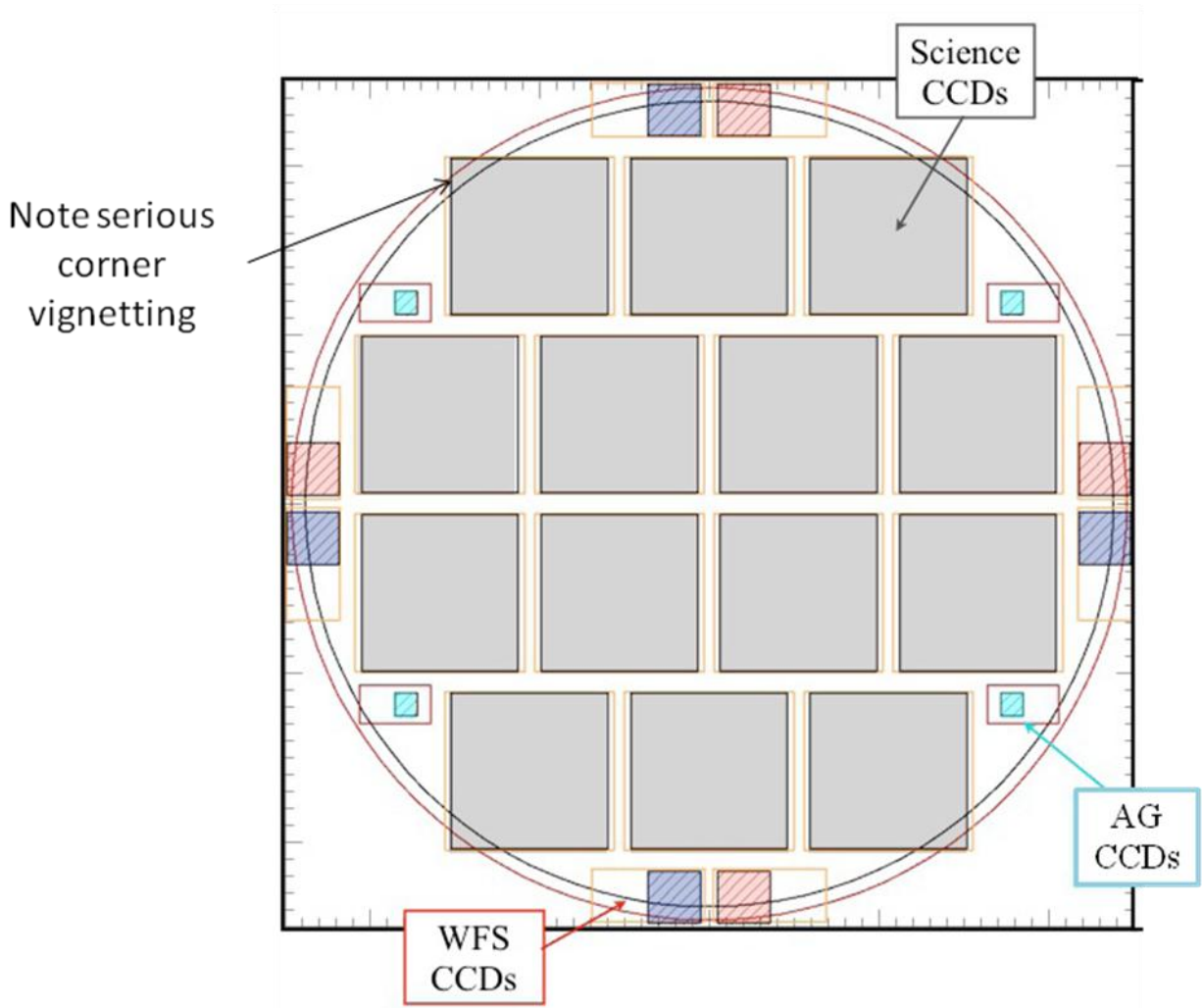

Figure 1. Conceptual layout of the JPCam detector focal plane prior to selecting the sensors. 14 full-frame, loosely packed, CCDs are shown populating the unvignetted FoV (black circle - the outer red circle represents the region of <10% vignetting). The small $1k^2$ CCDs (cyan) are guider chips while the larger $2k^2$ CCD pairs (blue/red) represent the intra- and extra-focal plane curvature sensors.

JPCam's control electronics is required to allow the user to modify the read-out time (up to a maximum pixel rate of ~2MHz) and binning factor while defining regions of interest. The selection will be done through low-level routines controlled from a PC. Priority will be given to minimising the read-out time, for a given read-out noise (RoN) level; to optimize this trade all 16 channels of the detector will be read out in parallel. The performance of the detector control electronics is required not to significantly degrade the intrinsic performance of the science detectors, although differential read circuitry is being carefully considered. After competitive analysis in light of JPCam's requirements, e2v was selected to produce not only the detectors but also the full cryogenic camera system including focal plane integration and camera electronics[8]. The FSU will be designed and constructed by a Brazilian consortium.

# 3. JPCAM STRUCTURAL ELEMENTS

Because of the T250 telescope's very wide FoV combined with the confirmed excellence of the OAJ's intrinsic site seeing, we are required to fully optimize and maintain the image quality across the full focal plane of the mosaic. Optical analysis reveals that it is necessary, not only to continuously adjust the hexapod supporting the secondary (M2) mirror of the telescope, but also to supply semi-continuous positional actuation of the focal plane itself. The four curvature wave-front sensors in the periphery of the FoV, sampled every few minutes, give the information needed to correct both M2 and the camera's focal plane, analysis of the motions of which have demonstrated the need for a further hexapod actuator system (HAS) operating at the camera/telescope interface. The HAS system will be supplied by the NTE-Sener group based in Barcelona, Spain.

Given the full-up ~650kg weight of the cryostat holding the CCD mosaic, together with its associated electronics, cooling and vacuum system, the HAS bridges directly between the cryostat support structure (CSS) and the telescope's instrument support structure (ISS) thus avoiding contact with the much lighter filter/shutter unit (FSU) whose flexure requirements are not at the level which would impact image quality. In this way we avoid making the FSU, which carries much of the moving mechanisms of the system, structurally supportive of the much heavier cryostat. The structural assembly of JPCam is given in Fig.2 which shows the linkages between its various elements.

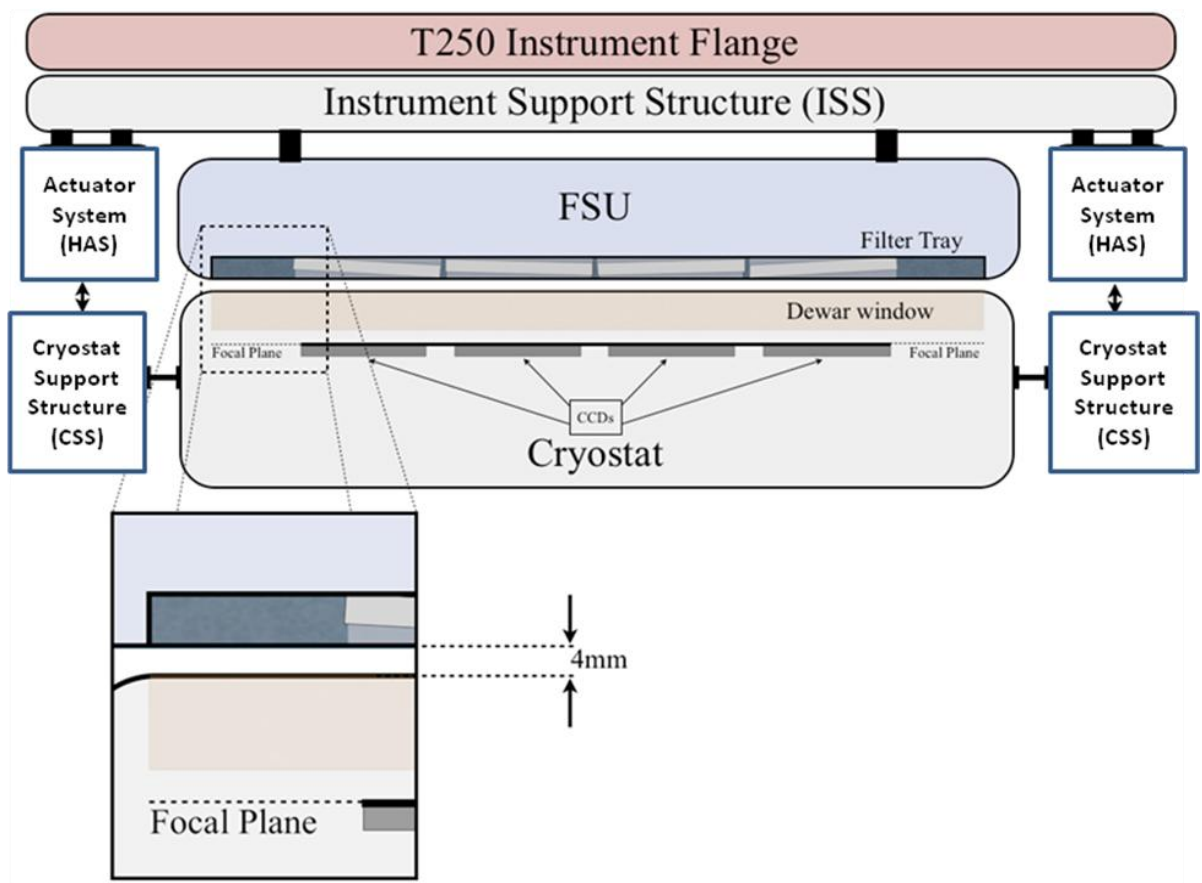

Figure 2. The structural elements of JPCam showing: the telescope interface flange directly mated with the ISS which supports the cryostat through linking the ISS with the CSS. Hexapod actuation of the cryostat with respect to the ISS is achieved through the HAS which bypasses the more delicate mechanisms associated with the FSU.

Shown in the expanded view of Fig.2 is the nominal, 4mm, gap between the FSU filters and the cryostat window. It is this gap which is continually modified by the actuation of the HAS which has a full piston range of ±2mm. Also to be noted is that the filters are tilted with respect to the optical axis of the system; this feature will be described in more detail in the following section.

## 4. THE FILTER/SHUTTER UNIT

The Filter/Shutter Unit (FSU) is designed to admit 5 filter "trays"; all five of which contain 14 square filters each corresponding to the 14 CCDs of the detector mosaic. Additionally, the filter trays also have filter holders for broad-band filtering of the 12 auxiliary wave-front sensing (8) and guider (4) chips. As shown in Fig.1, some degree of vignetting is inevitable given the size of the science CCDs. If the gap between CCDs is increased this then relaxes the requirement for close-packing of the filters however, this is at the expense of an increase in vignetting at the corners of the mosaic. On the other hand, if the CCD gap is decreased, then the foot-print of the filters on their corresponding detector will give more uniform edge vignetting around its periphery. The impact of such vignetting on the J-PAS survey efficiency was analysed in some detail and it was concluded to reduce the CCD gap so as to minimize corner vignetting while uniformly vignetting the edges of all CCDs. In this way, the 56-deep data-cube created at each sub-field of the survey will be uniformly vignetted throughout its depth. This is in preference to the alternative of having corner vignetting which, while effecting only a subset (16) of the 56 filter images, actually compromises all 56-deep pseudo-spectra corresponding to corners throughout the data-cube; the second alternative thus actually has a much more profound degradation of the survey as a whole. The two alternatives are depicted in Fig.3.

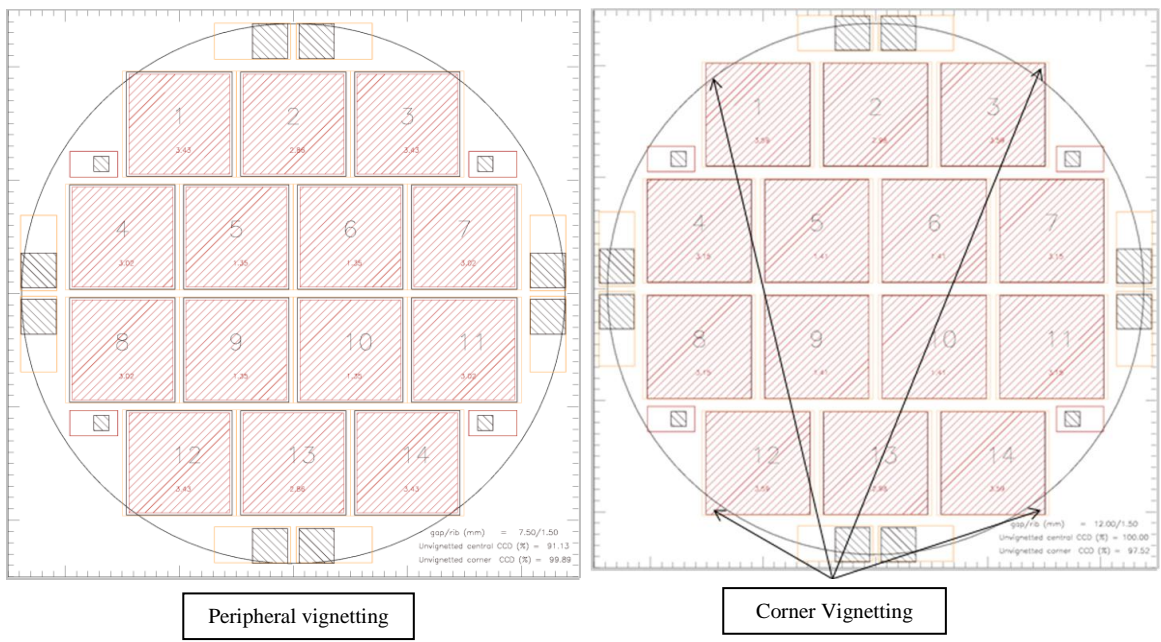

Figure 3. The two vignetting alternatives are shown. On the left-hand image the gap between detectors has been minimized which necessitates that the external filters are under-sized; this results in peripheral vignetting of each CCD. The alternative is shown in the right-hand figure whereby the size of the filters can now be increased to permit a fully unvignetted detector; in this case however, four corners of the mosaic (and the peripheral WFSs) are seriously vignetted.

The focal-plane of the T250 telescope is non-telecentric and hence, in order to retain the steepness of each intermediate-band filter waveband profile and the uniformity of its wavelength centering, the filter must be held in each tray so as to induce a differential tilt in each of the 14 filters of the mosaic, so that each filter is perpendicular to the chief ray at its centre. This amounts to a maximum tilt of ~3.5º for the outer filters of the mosaic equivalent to a ~6mm departure from a flat surface. Furthermore, in order to minimize the peripheral vignetting of the CCD by its corresponding filter, the distance between filters and CCDs is required to be as close as practical; as stated previously, a nominal gap of 4mm between the filters and the dewar window has been chosen to allow for filter tray deployment and the necessity of pistoning the mosaic focal plane with the HAS. In order to prevent frost and/or condensation from forming on the large (~500mm dia.) dewar window, the full system from the T250 telescope's wide-field corrector through the FSU to the cryostat (as actuated by the HAS) will be sealed and over-pressured with dry air or $N_2$.

The 5 filter trays are selectable remotely so the FSU will include the motors and encoders and the control system needed for their operation. Each filter tray is designed to be easily and manually removable and exchangeable from the closed frame. Individual filters can be manually removable from their tray once the tray has been removed from the module. Details of the FSU are shown in Fig.4 while the full system concept design, including the NTE-Sener actuator system and the e2v detector assembly is shown in Fig.5. It should be emphasised that all these drawings are at a pre-Concept Design Review phase and hence should be regarded as work-in-progress and subject to change.

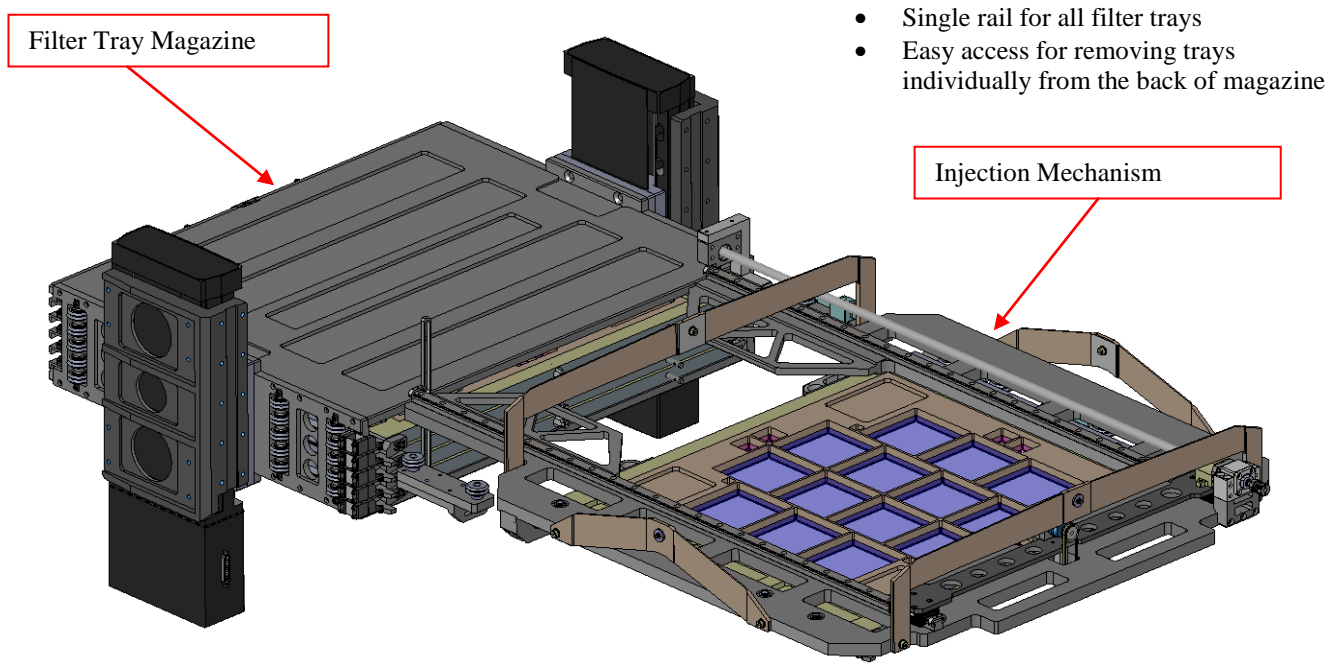

Figure 4. Four filter trays, each containing 14, 12.5nm bandwidth, survey filters and 12 broad-band auxiliary filters, are deployed above the focal plane using an exchange mechanism which selects the chosen filter tray from a magazine. Only the 4 filter trays, containing within them a total of 56 survey filters, are shown in the figure. Subsequent to this drawing, a requirement for an extra filter tray (5 in all) containing 14 identical broad-band R filters has been requested.

## 5. CRYOSTAT AND DETECTOR FOCAL PLANE

The cryostat and detector focal plane for JPCam is being supplied by e2v[8]. The layout of the focal plane cold plate (FPCP) is given in Fig.6 where the 14 science sensors and 12 auxiliary guide and WFSs are shown. As a risk mitigation strategy in the context of a 4-5 year continuous survey, the cryostat will be $LN_2$ cooled. An large $LN_2$ tank will auto-feed the cryostat through routing of the cooling lines via the telescope cable wraps and thence through two rotational couplers as required to accommodate both cassegrain and altitude rotation.

As stated previously, the cryostat is held directly to the instrument support structure (ISS) via the hexapod actuator system (HAS) which is required in order to fully optimize image quality across the full FoV. In Fig.7 we depict the curvature sensing method that will be adopted to calculate the Zernicke polynomials supplied to the M2 hexapod and the

HAS, while Fig.8 shows the expected improvement in imaging performance. The 4 pairs of intra- and extra-focal plane WFSs are offset from the focal plane by ±1mm; the $2k^2$ format of the WFS is sufficient to ensure the detection of several stars at the required SNR for each sensor every ~100s.

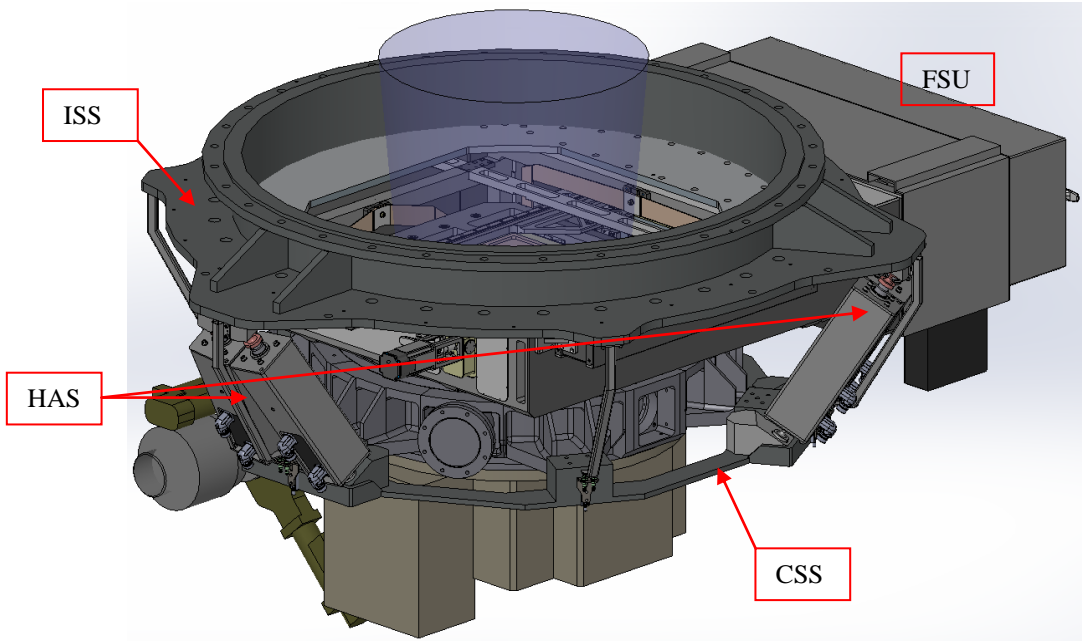

Figure 5. The FSU module, seen from the telescope side, as mounted on the instrument support structure (ISS) which itself bolts directly to the telescope's cassegrain instrument flange. The HAS actuator system is shown bridging the FSU spanning directly between the lower cryostat support structure (CSS) and the ISS.

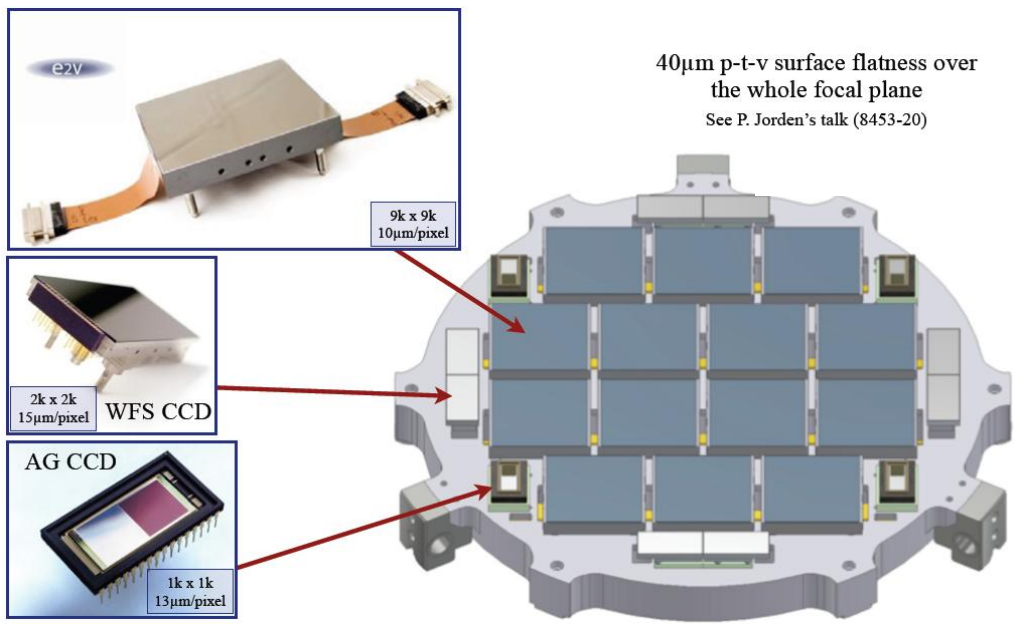

Figure 6. JPCam's focal plane layout as supplied by e2v. The 14 loosely packed, full-wafer, e2v science sensors are shown mounted on the FPCP. In the periphery are mounted 4, $1k^2$ frame-transfer (FT) guide CCDs and 4 pairs of $2k^2$ FT WFSs.

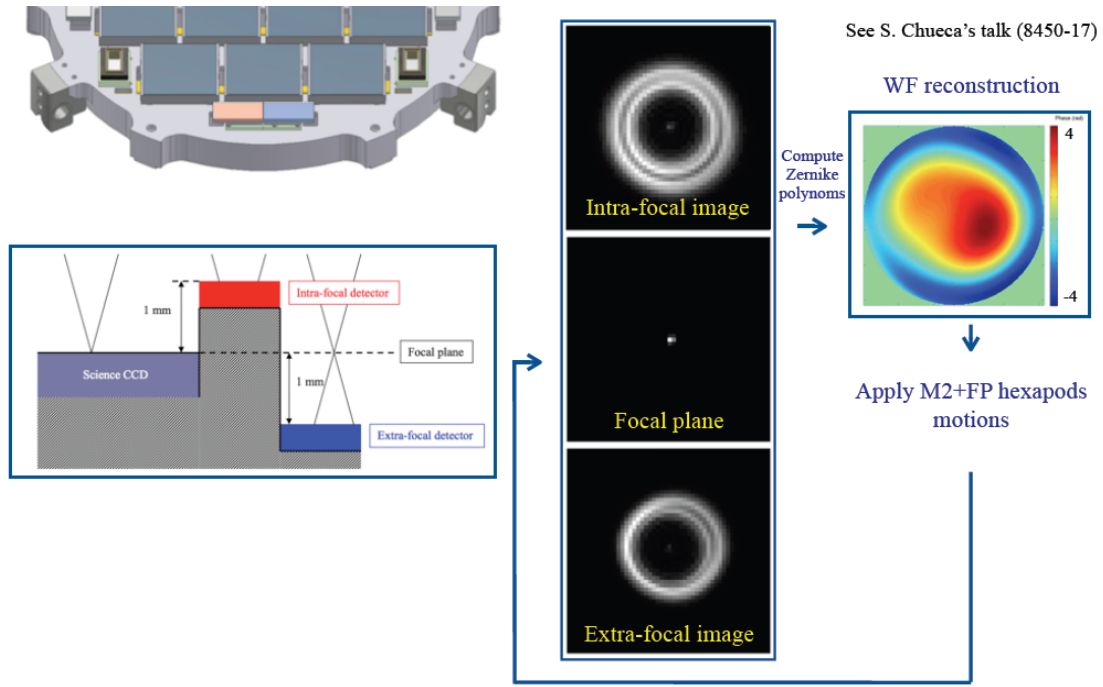

Figure 7. Lower right depicts the intra- and extra-focal plane WFSs giving rise to alternate out-of-focus images which can be analysed to compute Zernike polynomials which are used as inputs to the M2 hexapods and the HAS.

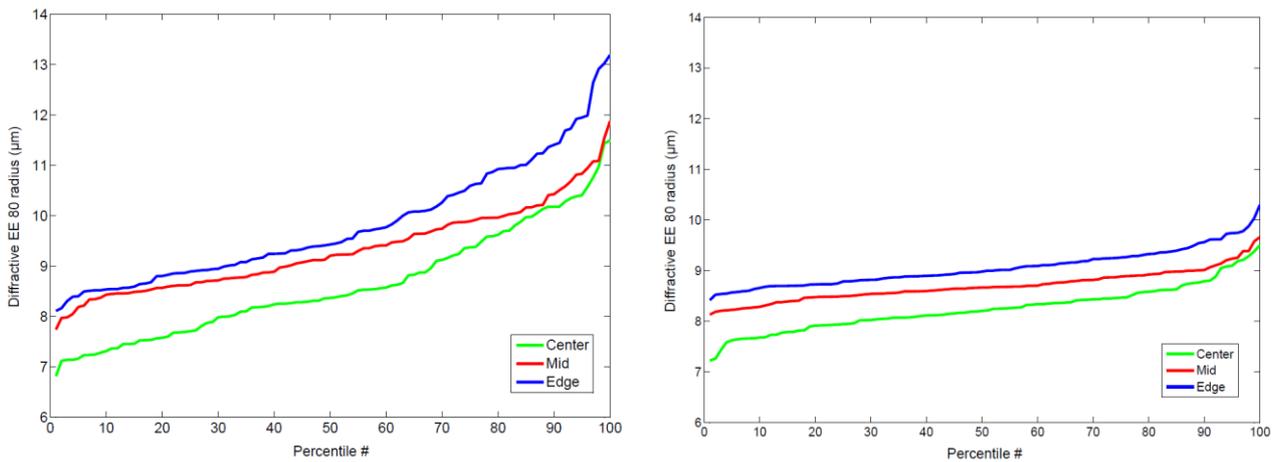

Figure 8. The telescope performance as determined through Monte Carlo simulations of the estimated sub-system tolerances as given by AMOS; the statistical distributions are shown for three positions across the FoV. The left-hand graph shows the result of M2 actuation alone, while the right-hand graph incorporates piston correction in the focal plane. Tip/tilt optimization is required in both cases.

# 6. J-PAS SCIENCE

J-PAS is a 8000 deg$^2$ photometric sky survey taylored for cosmology, that uses an unprecedented system of 54 narrow band filters, plus three broad-band U, R and Z filters, that will allow a redshift precision $\sigma_z$ ~0.003(1+z), sufficient to obtain a much better precision in the Dark Energy parameters than its predecessors. This precision in redshift can be achieved for bright, red galaxies, featuring a prominent 4000Å break, by using a filter system comprising 54 filters, each with a width close to 138Å, covering the wavelength range from ~3700 to ~9200Å, supplemented by 3 broad-band filters.

Based on the area the 8000 deg$^2$ we will target area and the time that we have to complete the main astrophysical survey (4-5 years, depending on the performance of the instrument), we estimate that we will be able to detect astronomical objects down to an apparent magnitude (3 arcsec. aperture) of AB ~22.5 – 23.5, depending on the wavelength, at 5$\sigma$ level. That includes billions of stars, hundreds of millions of galaxies, millions of quasars and AGNs, as well as thousands of variable objects such as supernovas and minor planets of the solar system. In terms of redshift, we will be able to observe galaxies up to z~1.3, emission-line galaxies up to z~2.5, quasars up to z~6, and supernovas up to z~0.6. In the standard cosmological model (the so called flat ΛCDM model), the distance to z=1 is 3,300 Mpc, or 10.8 billion light-years, and the universe at z=1 was only 6 billion years old (it is about 13.7 billion years old now).

J-PAS will be carried out with a telescope/camera combination with an etendue about 26m$^2$.deg$^2$, equivalent to a 2.5 meter telescope equipped with a 6deg$^2$ FoV camera, and covering 8000deg$^2$ in the sky in four years. We expect to measure positions and redshifts for over 14 million red, early-type galaxies with L > L* and $i_{AB}$ <22.5 in the redshift interval 0.1<z<1.0, with a precision $\sigma_z$ < 0.003(1+z). This population has a number density n >10$^{-3}$ Mpc$^{-3}$ h$^3$ galaxies within the 9 Gpc$^3$ h$^{-3}$ volume to be sampled by our survey, ensuring that the error in the determination of the BAO scale is not limited by shot noise. By itself, such a survey will deliver precisions of order <5% in the dark-energy equation of state parameter w, if assumed constant, and can determine its time derivative when combined with future cosmic microwave background measurements. In addition, J-PAS will yield high-quality redshift and low-resolution spectroscopy for hundreds of millions of other galaxies, including a very significant high-redshift population. The data set produced by this survey will have a unique legacy value, allowing a wide range of astrophysical studies.

The scientific program will have as central target the accelerating expansion of the Universe and the nature of its cause, the Dark Energy. The Collaboration plans to meet this challenge by measuring the dark energy equation of state parameter, w, primarily through the analysis of the Baryonic Acoustic Oscillations (BAO). Other complementary techniques will be also used in the same survey, including but not limited to: (i) the redshift distribution and clustering evolution of galaxy clusters, and (ii) weak gravitational lensing on large scales. All the efforts will also be done to implement the technique appropriate to measure distances with type Ia Supernovae.

The output of the survey will also enable addressing a wide variety of astrophysical problems, ranging from Galaxy Evolution to QSOs, exoplanets, Supernovae, asteroids, and different cosmological tests.

J-PAS essentially delivers ultra-low resolution spectroscopy of the whole northern sky using 56 wavelength samples equivalent to an integral field spectrograph (IFS) with R ~25 over >8000 deg$^2$. An early version of the filter distribution, overlaid upon which is a typical galaxy spectrum, is shown in Fig.9.

There is therefore an immense legacy value due to the nature of the survey. Studies of clusters and groups of galaxies, galaxy evolution, quasar studies, star formation, stellar astrophysics, halo stars, asteroids, weak lensing, and more become possible.

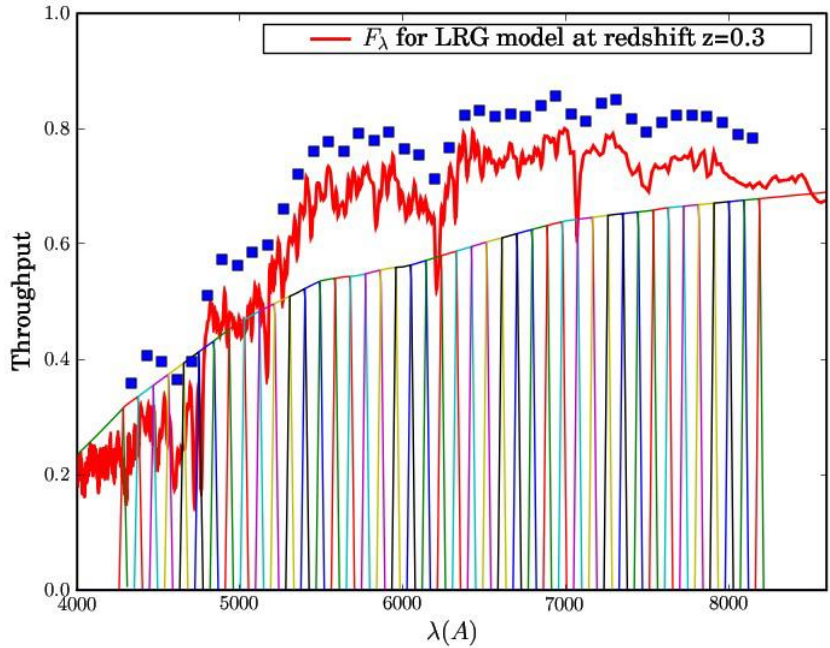

Figure 9. A JPCam IFS spectrum simulated by sampling a typical z=0.3 LRG. 42 of the 56 band-passes are displayed as a function of wavelength and instrumental efficiency. The blue dots represent flux measures obtain from the survey. Clearly ultra low-R, but nevertheless useful, spectra are recovered.

## 7. STATUS AND SCHEDULE

All elements of JPCam are nearing the end of their concept design phases.
- FSU – Brazilian consortium, based in Sao Jose dos Campos, SP, Brazil;
- Cryostat camera – e2v, Chelmsford, UK;
- HAS – NTE-Sener, Barcelona, Spain;
- Interface management – AMOS, Leige, Belgium.

These 4 entities will be convening later in July'12 to review progress and iron out critical interfaces. Once individual sub-assemblies have been through rigorous in-factory validation, we anticipate final assembly, integration and test at the T250 telescope to be scheduled to begin in Q2'14, with end of telescope commissioning and the beginning of J-PAS science sometime in Q3'14.

## 8. ACKNOWLEDGEMENTS

This project is funded by the Brazilian agencies CNPq (Conselho Nacional de Desenvolvimento Científico e Tecnológico), FAPESP (Fundação de Amparo à Pesquisa do Estado de São Paulo), FINEP (Financiadora de Estudos e Projetos) and FAPERJ (Fundação de Amparo à Pesquisa do Estado do Rio de Janeiro); and on the Spanish side through CEFCA (Centro de Estudios de Física del Cosmos de Aragón).

# REFERENCES

[1] Andrés Javier Cenarro et al., "The Javalambre Astrophysical Observatory project", Proc. SPIE 7738 (2010).
[2] Andrés Javier Cenarro et al., "The Observatorio Astrofisico de Javalambre: telescopes and instrumentation", Highlights of Spanish Astrophysics VI, p. 680-685 (2010).
[3] Andrés Javier Cenarro et al., "The Observatorio Astrofisico de Javalambre: goals and current status", Proc. SPIE 8448 (2012).
[4] Mariano Moles et al., "The Observatorio Astrofisico de Javalambre. A planned facility for large scale surveys", Highlights of Spanish Astrophysics VI, p. 73-79 (2010).
[5] http://j-pas.org
[6] Antonio Marin-Franch et al., "Design of the J-PAS and J-PLUS filter systems", Proc SPIE 8450 (2012)
[7] http://www.bonn-shutter.de/
[8] Paul Jorden et al., "A gigapixel commercially manufactured cryogenic camera for the J-PAS 2.5m Survey Telescope", Proc SPIE 8453 (2012)